\documentclass[pdflatex,sn-aps, iicol]{sn-jnl}%

\usepackage{upgreek}
\newcommand{\resim}{\mathord{\sim}}

\begin{document}

\title{LNoS: Lithium Niobate on Silicon Spatial Light Modulator}

\author*[1]{\fnm{Sivan} \sur{Trajtenberg-Mills}}\email{sivantra@mit.edu}
\equalcont{These authors contributed equally to this work.}

\author*[1]{\fnm{Mohamed } \sur{ElKabbash}}\email{melkabba@mit.edu}\equalcont{These authors contributed equally to this work.}

\author[1]{\fnm{Cole J.} \sur{Brabec}}

\author[1]{\fnm{Christopher L.} \sur{Panuski}}

\author[1]{\fnm{Ian} \sur{Christen}}

\author[1]{\fnm{Dirk} \sur{Englund}}\email{englund@mit.edu}

\affil*[1]{\orgdiv{Research Laboratory of Electronics}, \orgname{ Massachusetts Institute of Technology}, \orgaddress{\street{50 Vassar St}, \city{Cambridge}, \postcode{02139}, \state{MA}, \country{USA}}}

\abstract{Programmable spatiotemporal control of light is crucial for advancements in optical communications, imaging, and quantum technologies. Commercial spatial light modulators (SLMs) typically have megapixel-scale apertures but are limited to $\sim$kHz operational speeds. Developing a device that controls a similar number of spatial modes at high speeds could potentially transform fields such as imaging through scattering media, quantum computing with cold atoms and ions, and high-speed machine vision, but to date remains an open challenge. In this work we introduce and demonstrate a free-form, resonant electro-optic (EO) modulator with megapixel apertures using CMOS integration. The optical layer features a Lithium Niobate (LN) thin-film integrated with a photonic crystal (PhC), yielding a guided mode resonance (GMR) with a $Q$-factor $>$ 1000, a field overlap coefficient $\sim$90$\%$ and a 1.6 GHz 3-dB modulation bandwidth (detector limited). To realize a free-form and scalable SLM, we fabricate the PhC via interference lithography and develop a procedure to bond the device to a megapixel CMOS backplane. We identify limitations in existing EO materials and CMOS backplanes that must be overcome to simultaneously achieve megapixel-scale, GHz-rate operation. The ``LN on Silicon'' (LNoS) architecture we present is a blueprint towards realizing such devices.}




\maketitle

\section{Introduction}\label{sec1}

Spatial Light Modulators (SLMs), optoelectronic devices that modulate the amplitude, phase, or polarization of light, have played a pivotal role in a wide range of applications including optical communications \cite{Communication_SLM}, commercial displays, beam shaping and steering \cite{laser_Machine_SLM, Libster-Hershko:15}, photolithography \cite{del2019light}, augmented and virtual reality \cite{ARVR}, holography \cite{20years, tsang2016review}, structured light \cite{Structuredlight_SLM}, super-resolution imaging \cite{Superesolution_SLM}, machine learning \cite{Liane_SLM}, astronomy \cite{gardner2006james}, and more. Recently, SLMs were instrumental in quantum simulators, facilitating the preparation of two-dimensional neutral atom arrays \cite{QuantumSLM_ebadi} for programmable simulation of the quantum spin model \cite{QuantumSLM_Giulia}. 

\begin{figure*}
\centering
\includegraphics[width=\textwidth]{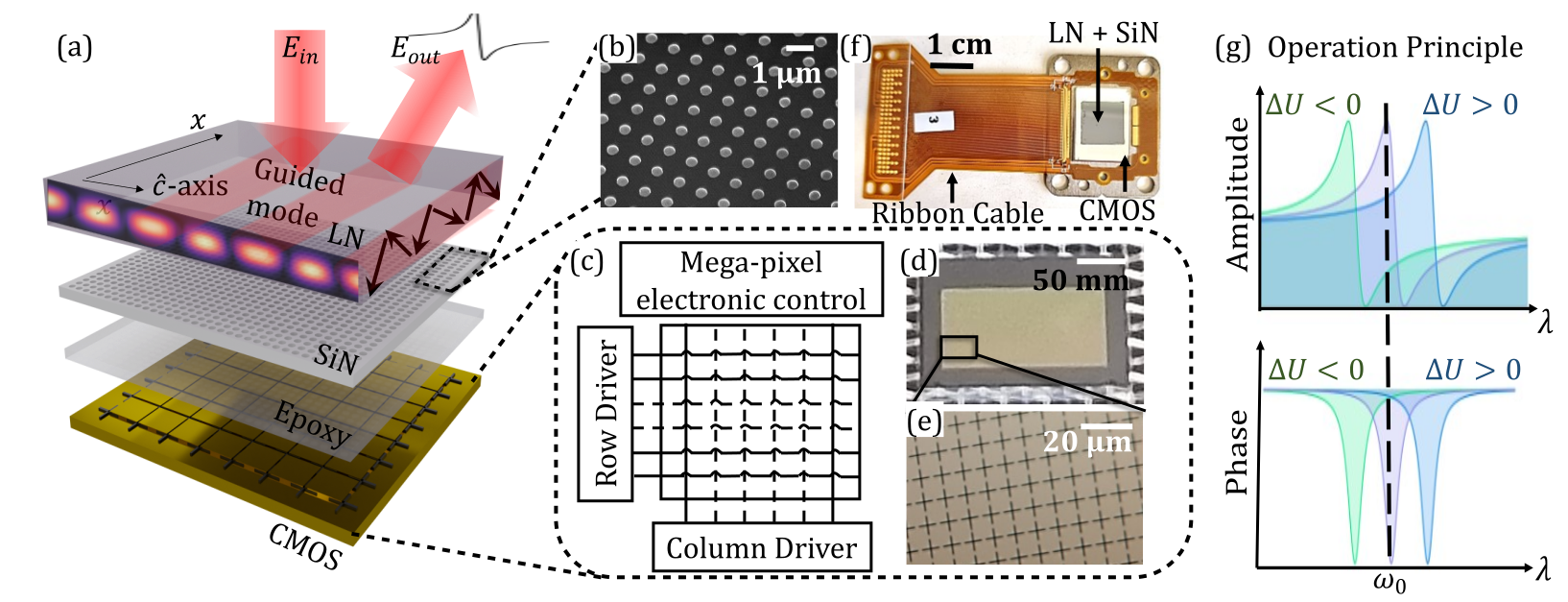}  \label{fig:concept}
\caption{\textbf{Lithium Niobate on Silicon (LNoS) SLM.} (a) Schematic depiction of the CMOS-integrated architecture. A SiN PhC excites a guided mode in the LN film, enabling free-form control via an epoxy-bonded CMOS backplane. (b) A representative scanning electron microscope (SEM) image of the fabricated PhC. (c) Megapixel control is achieved using row-column addressing. (d/e) Microscopic images of the commercial CMOS backplane. (f) The final fabricated device. (g) Operation principle: a voltage ($\Delta U$)-induced perturbation in the LN refractive index shifts the resonance, changing the reflected phase and amplitude at the center frequency $\omega_0$.}
\end{figure*}

However, commercially-available SLMs typically operate at or below kHz-order modulation rates \cite{LSA_LCOS, elkabbash2023metal}. GHz-rate modulation bandwidths are necessary for a wide range of applications including optical communication, video holography \cite{20years}, imaging through dynamic scattering media \cite{imagingScattering_roadmap}, and ultrafast laser pulse shaping \cite{capasso_SLM}. Recently, nanophotonic resonances modulated through EO effects (e.g., Pockels, Kerr \cite{capasso_SLM, LN16, programmablePlasmonic_SLM}, or free-carrier plasma dispersion \cite{Acs_ITO, Nanotech_ITO, Chris}) have achieved GHz-rate modulation bandwidth, but with limited ($<$100) pixel counts.

One common factor limiting pixel count is the planar interconnect architecture, which requires controls for the two-dimensional (2D) modulator array to be routed through a limited perimeter length (for wires of finite size with finites distance between them). For example, a sample device with a 100 $\upmu$m wire spacing and a 1 cm$^2$ square aperture is limited to $\resim$400 devices. Similar limitations are present in silicon optical phased arrays \cite{OpticalPhaseArray_Nature2013}, especially considering the size constraints of efficient free-space couplers. Scaling to megapixel-scale 2D apertures has only been achieved with CMOS backplanes; however, to date, no work has combined this commercial architecture with high-speed EO modulation. 

Here, we introduce such a device by coupling a EO-modulated LN guided mode resonance (GMR) to a CMOS backplane (Figure \ref{fig:concept}(a)). Tuning the LN film's refractive index through the Pockels effect controls the amplitude and phase of reflected light. We optimized tuning efficiency by maximizing the optical field overlap with the LN film ($> 90\%$) and show that the 3-dB modulation bandwidth reaches 1.6 GHz, limited by our detector's bandwidth. Furthermore, we demonstrate beamforming with a few-pixel prototype. Finally, we scale the design through: 1) interference lithography-based fabrication of the guided resonance's photonic crystal (PhC; Figure \ref{fig:concept}(b)) and 2)  bonding to a commercial, megapixel CMOS backplane (Figure \ref{fig:concept}(c-e)). Our SLM retains the main advantages of commercial devices -- namely, the use of a CMOS backplane and free-form refractive index patterning -- while removing the mechanical bandwidth limitations associated with liquid crystal on silicon (LCoS)- and micro-electro-mecanical system (MEMS)-driven devices. These results allow us to explore the limitations of current EO materials and electronic backplanes that must be addressed to realize GHz-rate, megapixel-scale SLMs.

\section{Results}\label{sec2}

\subsection{Operation Principle}\label{operation}

The CMOS backplanes characteristic of LCoS (Liquid Crystal on Silicon) SLMs offer two unique features that enable megapixel apertures: 1) multilayer routing and control electronics beneath the pixel arrays to maximize pixel density, and 2) free-form refractive index modulation, which eliminates the need to align the photonic (LC molecules) and electronic (CMOS backplane pixels) components. Combining this commercial architecture with high-speed EO modulation enables us to achieve free-form, large-scale, GHz-rate spatial light modulation.

However, high-speed EO effects, such as the Pockels or Kerr effects, are limited by the relatively small changes they induce in the refractive index. A backplane-induced electric field $E$ along the $c$-axis in the EO film produces a co-aligned refractive index change
\begin{equation}
\label{eq:delta_n}
\Delta n = \frac{n^3}{2}r_{33} E 
\end{equation}
for the EO coefficient $r_{33}$. LN exhibits a relatively high $r_{33}\sim$30 pm/V. In our architecture, the field $E$ affecting the LN layer is the fringe field between neighbouring pixels. An approximate upper limit for the fringe field is the electric field of a parallel plate capacitor $E=V/L$ for a voltage $V$ and pixel spacing $L$. Given typical CMOS values ($V<5 \text{V}$) and $L \approx 5$ $\upmu$m), Equation \ref{eq:delta_n} yields $\Delta n \approx 10^{-4}$. This means that the phase shift for an LN film with thickness of $500$ nm at wavelength $\lambda = 1550$ nm is $\approx 0.002$ rad, insufficient for most practical applications.

To amplify the index modulation, we introduce a locally-modulated GMR. Figure \ref{fig:concept}(g) shows how by applying an electric field and thus changing the LN refractive index, effectively introducing the resonator to and increase in the propagation length to $Q/k$, and thus shifting the resonance and amplifying the phase-amplitude response. From perturbation theory \cite{joannopoulos2008molding}, the shift in resonant wavelength
\begin{equation}\label{eq:delta_lambda}
\frac{\Delta\lambda}{ \Gamma} = \frac{\Delta n}{n}\eta Q
\end{equation}
relative to the linewidth $\Gamma$ depends on the field overlap coefficient $\eta$ and quality factor $Q$. Several high-$Q$ resonances are compatible with the proposed design including resonances excited in high contrast grating modes \cite{Connie_HighContrastGratings}, bound-state in continuum modes \cite{ndao_BIC}, and surface lattice resonances \cite{SurfaceLatticeResonance}. We ultimately chose a GMR \cite{GMR} featuring a deposited PhC. A distinct advantage of the GMR is the tunable $Q$-factor, which is a function of the hole size in the 2D PhC that couples bound slab modes to the radiative continuum. Additionally, the large-area resonance \cite{nonlocal} enables free-form patterning without precise alignment between the PhC-LN device and the CMOS backplane. Moreover, our approach avoids the often demanding nano-patterning of LN \cite{LN_fab} and increases the field overlap coefficient $\eta$. Our optimized GMR achieves $Q\approx 1,000$ and $\eta \approx 90\%$ to maximize tuning efficiency. Consequently, a small local index perturbation significantly modulates a reflected beam's amplitude and phase. We discuss the field overlap optimization in the following section. Figure \ref{fig:concept}(f) shows the final fabricated device, where the entire stack is wire bonded to a ribbon cable for row-column addressing through a micro-controller. 

\begin{figure*}
\centering
\includegraphics[width=\textwidth]{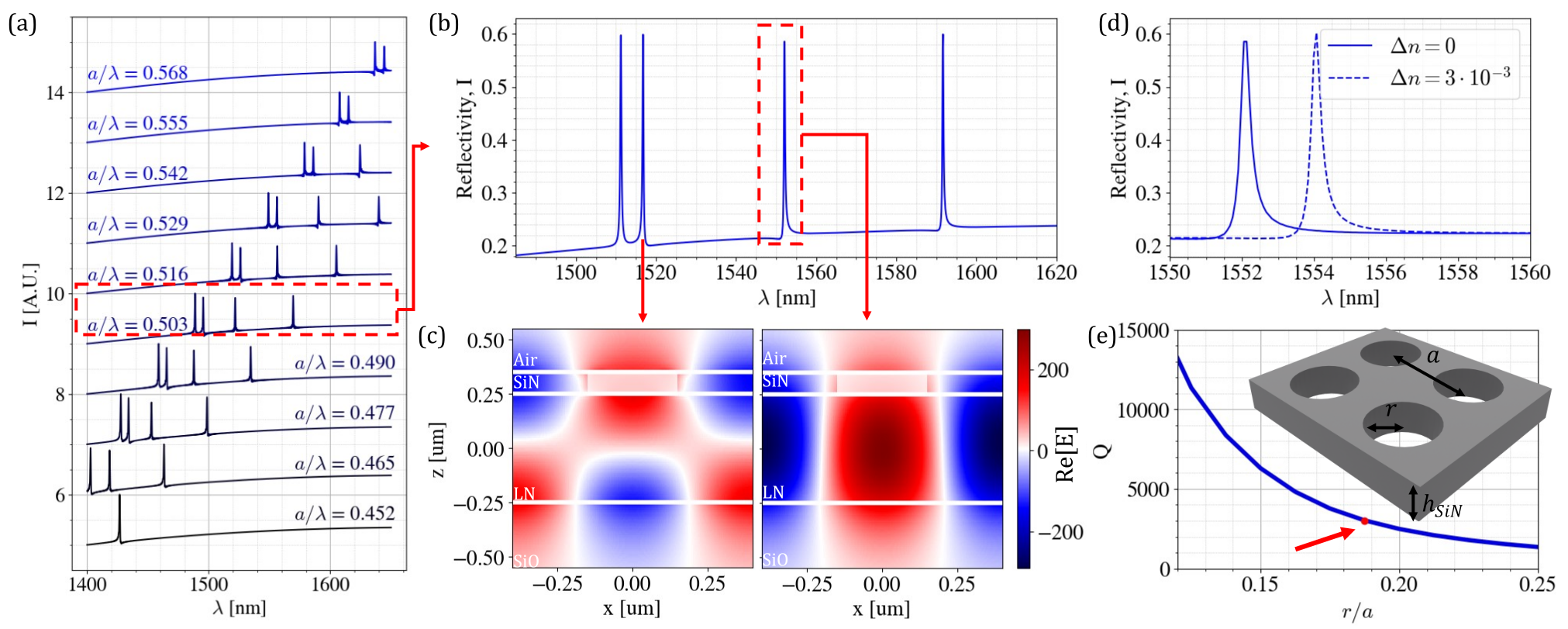}
\caption{\textbf{Optimized LN GMR Design.} (a) Simulated reflectance $I$ spectrum for different wavelength-normalized lattice constants $a/\lambda$ of the square hole array. (b) Simulated reflected spectrum for $a = 800$ nm, showing a clear resonance at the $\lambda_0 = 1552$ nm operating wavelength. (c) Finite difference time domain (FDTD) simulated optical field at resonance, for the low-order mode at $\lambda = 1552$ nm (right) and higher-order mode  $\lambda = 1516$ nm (left) respectively. The calculated field overlap coefficient at $\lambda_0$ is $\eta \approx 89\%$. (d) Modulation via local-index perturbation, illustrated by the simulated reflectance spectrum with (blue, dashed line) and without (blue, full line) a $\Delta n =3 \times 10^{-3}$ index perturbation. (e) $Q$ factor vs. normalized hole radius $r/a$. Our chosen working point $r/a\approx 0.19$ (red dot) offers $Q \approx 1,000$.}
\label{fig:design}
\end{figure*}

\subsection{LN GMR modulator design}
\label{Design}

The LN GMR design consists of a 500 nm-thick LN film where light is confined in a guided mode. The LN thickness is chosen to maximize internal field confinement at the operational wavelength, thereby maximizing tuning efficiency. On top of the LN, we deposit a layer of a 100 nm-thick Silicon Nitride (SiN), with a PhC consisting of a square lattice of holes (radius $r$, pitch $a$) etched into it.  The grating couples light into and out of the LN waveguide; its thickness was chosen to enable high $Q$-factor resonances for hole dimensions compatible with interference lithography as we will detail later. Figure \ref{fig:design}(a) shows the simulated reflectance versus PhC lattice constant $a$ for a normally incident plane wave polarized 45$^{\circ}$ with respect to the co-aligned PhC basis and LN $c$-axis. The same arrangement is used for our cross-polarized experimental setup described in Sec. ~\ref{subsubsec2_Charachterization}. The LN birefringence splits degenerate modes of the 2D GMR (i.e., along each grating axis), enabling both modes to be observed at different frequencies. For $ a/\lambda \approx 0.516$, choosing $a = 800$ nm yields $\lambda_0 \approx 1550$ nm (see Figure \ref{fig:design}a/b). This resonant mode is of interest because (i) it is polarized along the $c$-axis, and (ii) it corresponds to the zeroth order mode. Using the zeroth order mode maximizes the field overlap coefficient $\eta$, allowing for a larger response according to Equation \ref{eq:delta_lambda}. Figure \ref{fig:design}(c) compares the electric field magnitude for the zeroth order mode at 1550 nm (right) and first order mode at $1521$ nm (left; see Methods). The zeroth-order mode maximizes field confinement in the LN. In order to estimate $\eta$, we perturb the refractive index of the LN film by $\Delta n = 3 \cdot 10^{-3}$. The resulting simulated resonance shift is seen in Figure \ref{fig:design}(d). Using this shift, we calculate $\eta$\ from Equation \ref{eq:delta_lambda} and obtain $\eta \approx 89\%$. In comparison, the field overlap coefficient of a Mie resonant bound state in a continuum EO modulator is $\sim 25\%$ \cite{capasso_Mie}. 

Several design parameters can be used to tune the $Q$-factor of GMRs including the grating thickness, grating-waveguide index contrast, grating-waveguide separation, and the hole radius (in case of a 2D grating). Figure \ref{fig:design}(e) estimates the GMR $Q$ versus $r/a$ based on guided mode expansion \cite{Minkov2020}. We chose $r = 150$ nm, corresponding to $Q \sim 1,000$, to ensure stability against thermal/mechanical fluctuations and comparability with scalable fabrication techniques (where $r/a<0.25$ becomes challenging).

\subsection{Experimental Characterization of a 1D Modulator }\label{subsubsec2_Charachterization}

\begin{figure*}
  \centering \includegraphics[width=\textwidth]{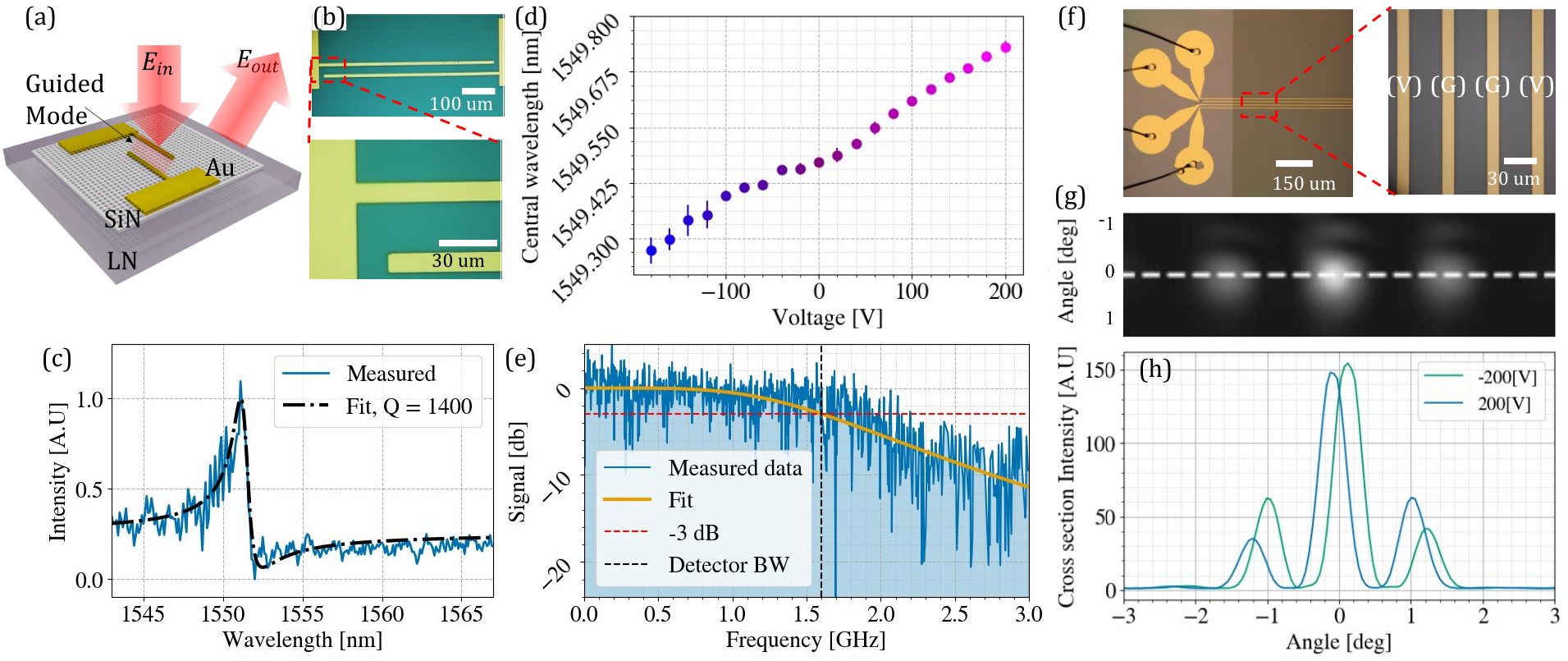}
\caption{\textbf{Characterization of the LN GMR modulator}. (a) Schematic representation of the device. (b) Micrographs of the fabricated device. (c) Experimentally measured spectrum of the GMR resonance (blue). The $Q$-factor $Q \approx 1400$ is extracted from the fit to a Fano lineshape (black dot-dash line). (d) Resonant wavelength shift as a function of the applied DC voltage. (e) Bandwidth measurement. Blue line shows the measured data while modulated with $V_{rms}$ = 2 V. Yellow line shows the fit to a second order frequency response. Red dotted line shows the extracted 3 dB bandwidth to be exactly 1.6 GHz, equal to our photo-detector bandwidth (black). (f) Images of the 3 pixel fabricated device with patterned gold electrodes. (g) Far-field image of the SLM as obtained on the camera. The white dotted line denotes the cross-section. (h) Cross section of the far field intensity near resonance (1547 nm), for different applied voltages. The far-field pattern is modulated due to the change in phase and amplitude of the SLM.}
  \label{fig:1D}
\end{figure*}

Before addressing the complications arising from a mega-pixel aperture, we first fabricated a single-pixel device with patterned electrodes (Figure \ref{fig:1D}a). After SiN thin film deposition, we patterned the PhC grating and 300 nm-thick gold electrodes (30 $\upmu$m gap) via electron-beam lithography (see Methods). Figure \ref{fig:1D}(b) shows images of the fabricated device. We used a cross-polarization reflection microscopy setup to isolate the GMR emission from resonant wavelength versus bias voltage (Figure \ref{fig:1D}d). The dominant linear trend observed is characteristic of the Pockels effect and outrules thermal effects (symmetric under polarity reversal). The small non-linearity is likely due to poling effects, which have been observed in LN at comparable bias levels \cite{Hu:21}. A $0.5$ nm maximum shift is obtained for a 400 V bias swing.

To determine $\eta$ experimentally, we first evaluate the DC field inside the LN film using COMSOL\textsuperscript{\textregistered} from which we calculate $\Delta n$ through Equation \ref{eq:delta_n}. Then using Equation \ref{eq:delta_lambda} we determine the experimental value of $\eta$ to be $\eta \approx 90 \%$ in close agreement with the simulations in Figure \ref{fig:design}(c). 
In order to measure the bandwidth of our device, we used an optical analyzer to measure the peak modulation amplitude at frequencies up to 2.5 GHz, as well as a lock-in measurement of the small amplitude response at high frequency. Figure \ref{fig:1D}(e) illustrates the peak EO modulation amplitude obtained from the optical analyzer with an input of 2 V. In order to overcome the noise arising from measurement instability, we fit a second-order frequency response (yellow, this is the expected response of a feedback TIA \cite{Razavi:12}) and find a 1.6 GHz 3-dB modulation bandwidth, equal to that of our photodetector (Methods).

To characterize the LN-GMR's beamforming capability, we fabricated a three-pixel electrode array (30 $\upmu$m pitch). Figure \ref{fig:1D}(f) shows the fabricated device wirebonded to an RF chip carrier for high-speed operation. This device can control the 1D wavefront in the far field by changing the applied voltage on the electrodes for a given wavelength. Figure \ref{fig:1D}(g-h) show the obtained image and a cross-section of the far-field intensity, at 1547 nm. A comparison with theoretical analysis of the beamforming capabilities of the device are presented in Appendix \ref{secA1}.

\begin{figure*}
  \centering
  \includegraphics[width=160mm]{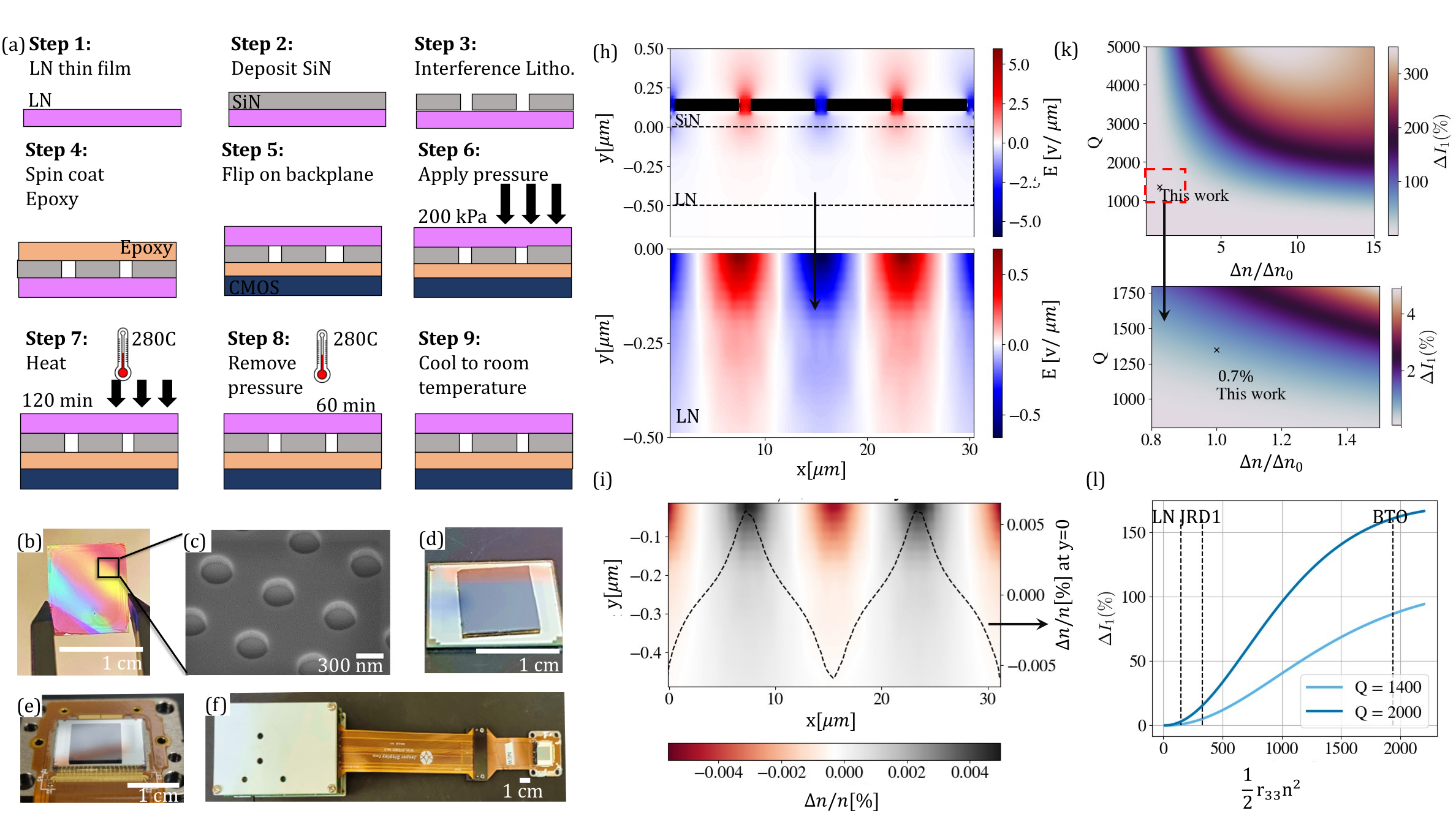} 
    \caption{\textbf{Scaling to mega-pixel apertures.} (a) Backplane bonding steps. (b/c) Sample images after interference lithography (Step 3). (d/f) Images of sample after backplane bonding and wirebonding, respectively. (f) Final device with microcontroler. (h) Simulated electric field $E$ with a periodic +5V and -5V applied on the electrodes (black). A magnified view of $E$ and the resulting change (in percent) in refractive index $\Delta n /n$ inside the LN layer (i) are shown below. The black dotted line in (i) shows $\Delta n /n$ at the LN-SiN interface. (k) The change (in percent) in the first diffraction order's intensity, $\Delta I_1$, for varying $Q$-factor and $\Delta n/ \Delta n_0$, where $\Delta n_0$ is the refractive index change achieved using our experimental parameters. $\Delta n/ \Delta n_0$ encompasses refractive index $n$, EO coefficient $r_{33}$, $\eta$, and electric field strength changes. Our work has $Q \approx 1400$ and is marked by a black $\times$. A zoomed-in view around our working point is shown below. (l) $\Delta I_1$ for a $5V$ peak bias voltage, $Q = 1400$ (light blue), and $Q = 2000$ (dark blue) as a function of the modified EO coefficient $\frac{1}{2}r_{33}n^2$ for LN (this work), JRD1, and BTO.}
  \label{fig:last}
\end{figure*}

\subsection{LNoS SLM: scaling to megapixel count}\label{subsec2}

After demonstrating GHz-rate modulation, we investigated scaling to megapixel apertures by integrating the LN GMR with a commercial CMOS backplane. This use of a well-established commercial technology allow us to scale pixel count while retaining fast EO modulation, but poses several challenges. Since the backplane area is approximately 1 cm$^2$, the PhC must be manufactured on a similar scale. However, electron beam lithography (EBL) is expensive and time consuming (on the order of 1 day/cm$^2$ for write time alone). To overcome this challenge, we fabricated the 2D PhC using interference lithography, which allows for efficient (minute-order exposures in our case), large-scale (cm$^2$) fabrication \cite{zhu2023metasurface}. Figure \ref{fig:last}(b/c) demonstrates the quality of the large-area interference lithography.

Direct integration of the LN GMR and backplane presents another challenge. A large-area, high-strength bond is desired; however, the fabrication process should not damage the LN GMR or modify its design parameters. Figure \ref{fig:last}(a) illustrates the custom bonding procedure we developed using a JD2552 2K voltage-drive backplane from Jasper Display Corp. We adapted a cold bonding method \cite{coldbonding} to adhere the GMR sample to the CMOS backplane. Following fabrication of the passive LN GMR samples, dies are spin coated with epoxy, flipped, and compression bonded to the backplane (see Methods for additional details). A 400 nm epoxy thickness minimizes backplane-LN separation (to optimize tuning efficiency) without inducing excess overlap between the GMR mode and aluminum electrodes. The distributed, non-local GMR makes the nanophotonic-electronic alignment tolerant to translational shifts, an advantage for this technique.  Figure \ref{fig:last}(d) shows the bonded sample. The bonding was robust to changes in temperature of up to 90 $^\circ$C, and to mechanical shock during experimentation. This sample was then wire bonded to a chip carrier (Figure \ref{fig:last}(e)), connected to a micro-controller as shown in Figure \ref{fig:last}(f), and temperature controlled via an attached Peltier module. While CMOS backplanes have been previously integrated into experimental SLMs \cite{zhu2023metasurface}, to our knowledge, this fabrication is the first demonstration of EO thin film bonding to a commercial backplane.  

To investigate the expected performance of the device and platform, and to find the ideal conditions under which LNoS can operate effectively, we evaluated the intensity change of the first diffraction order for a grating with double the pitch of the backplane. We simulated the spatial profile of the E field and associated refractive index change for the LNoS stack, shown in Figure \ref{fig:last}(h-i). Note that here we are considering the case of direct bonding of the SiN to the backplane without an epoxy \cite{reck2011fusion}. For simplicity, we choose the cross section at $y = 0$ $\upmu$m, where the field is strongest, to get an upper bound on the modulation efficiency. We then vary the GMR's $Q$-factor and $\Delta n / \Delta n_0$, the EO material's refractive index change relative to that of our experiment ($\Delta n_0$). This approach consolidates all factors affecting the wavelength shift --- including changes in refractive index $n$, EO coefficient $r_{33}$ and the electric field $E$ --- into a single parameter. Figure \ref{fig:last}(k) illustrates the change in the first-order intensity $\Delta I_1$, with our work denoted by a black $\times$. We confirmed this calculation using two different theoretical frameworks: Fourier analysis and coupled mode theory (see Methods). The anticipated $\Delta I_1$ is conservatively bounded at approximately 0.7$\%$ (confirmed by both methods). A similar analysis including the epoxy layer is included in Appendix \ref{secA3}. When performing the same calculation for the entire stack, as shown in Appendix \ref{secA3} in Figure \ref{fig:withepoxy} including the 400 nm of epoxy, this value drop to $\sim 0.09\%$. While thermal modulation was noted (due to local backplane heating), we were unable to detect these small intensity modulations experimentally.  

In the future, stronger EO materials could overcome the limited $\Delta n_0$ of our experiment. Figure \ref{fig:last}(m) depicts cross-sections of Figure \ref{fig:last}(l), i.e. the change in the first diffraction order intensity $\Delta I_1$ versus $\Delta n/\Delta n_0$. We consider the $Q$-factor used in our experiment ($Q = 1400$, light blue) as well as a higher $Q = 2000$ (blue, readily achievable by reducing hole diameter to 345 nm as shown in Figure \ref{fig:design}(e)). We selected a few popular EO materials for the comparison: JRD1 --- an EO polymer renowned for its high $r_{33}$ value of 100 pm/V and a refractive index of $n = 1.81$ \cite{Heni:17} --- and barium titanate (BTO), which affords $r_{33}\approx 342$ pm/V and a high refractive index of $n = 2.38$ \cite{abel2019large}. While LN and JRD1 exhibit similar performance, BTO would offer a significant advantage, potentially enabling a modulation of the first order by approximately 80$\%$. Additionally, increasing $Q$ to 2000, for instance, will almost double the modulation intensity. These results, alongside the newly developed fabrication method presented here, offer a roadmap for efficient high speed modulation with millions of pixels.  

\section{Conclusions}\label{sec13}

In conclusion, we have demonstrated a GHz-speed EO SLM architecture that is scalable to millions of pixels, bridging the gap between existing SLMs and high-speed optical modulators. By employing a CMOS backplane to pattern spatially varying electric fields within a high-$Q$ LN thin film GMR, we successfully demonstrate modulation bandwidths beyond 1.5 GHz. The scalability of the device was proven through the large area fabrication of a PhC using interference lithography and bonding the LN device to a commercial CMOS backplane. Furthermore, we explored the performance with different EO materials. The results indicate that material and resonator optimization could significantly improve modulation strength, opening up possibilities for a wide range of applications in classical and quantum computing, configurable optical add-drop multiplexing, AR/VR applications, LiDAR devices, video holography, imaging through scattering media, deep tissue photothermal therapy, laser cooling, to name a few. 

\backmatter

\section{Methods}\label{subsec4}
\subsection{Fabrication}\label{subsubsec4_fab}

\subsubsection{Electron Beam Lithography}
We used an X-cut LNOI (Lithium Niobate on Insulator) from NanoLN with a 500 nm-thick LN thin film, a 2400 nm buried oxide, and a 0.5 mm-thick silicon substrate. A 100 nm SiN film was deposited on the LN film via an STS PECVD system. The film was spin-coated with ZEP520A electron-beam resist at 4000 rpm and baked at 180$^\circ$C for two minutes. The hole array was patterned using an EBL-Elionix system with a 10 nA beam current and 350 $\upmu$C dosage. Development involved cold processing in Oxeylene at 0$^\circ$C for 75 s, followed by a 30 s IPA rinse. Pattern transfer to the SiN film was achieved using reactive ion etching (RIE) with CF$_4$ and O$_2$; resist was removed via a 24 hour NMP wet etch.

\subsubsection{Interference Lithography}
\begin{figure}
  \centering
  \includegraphics[width=74mm]{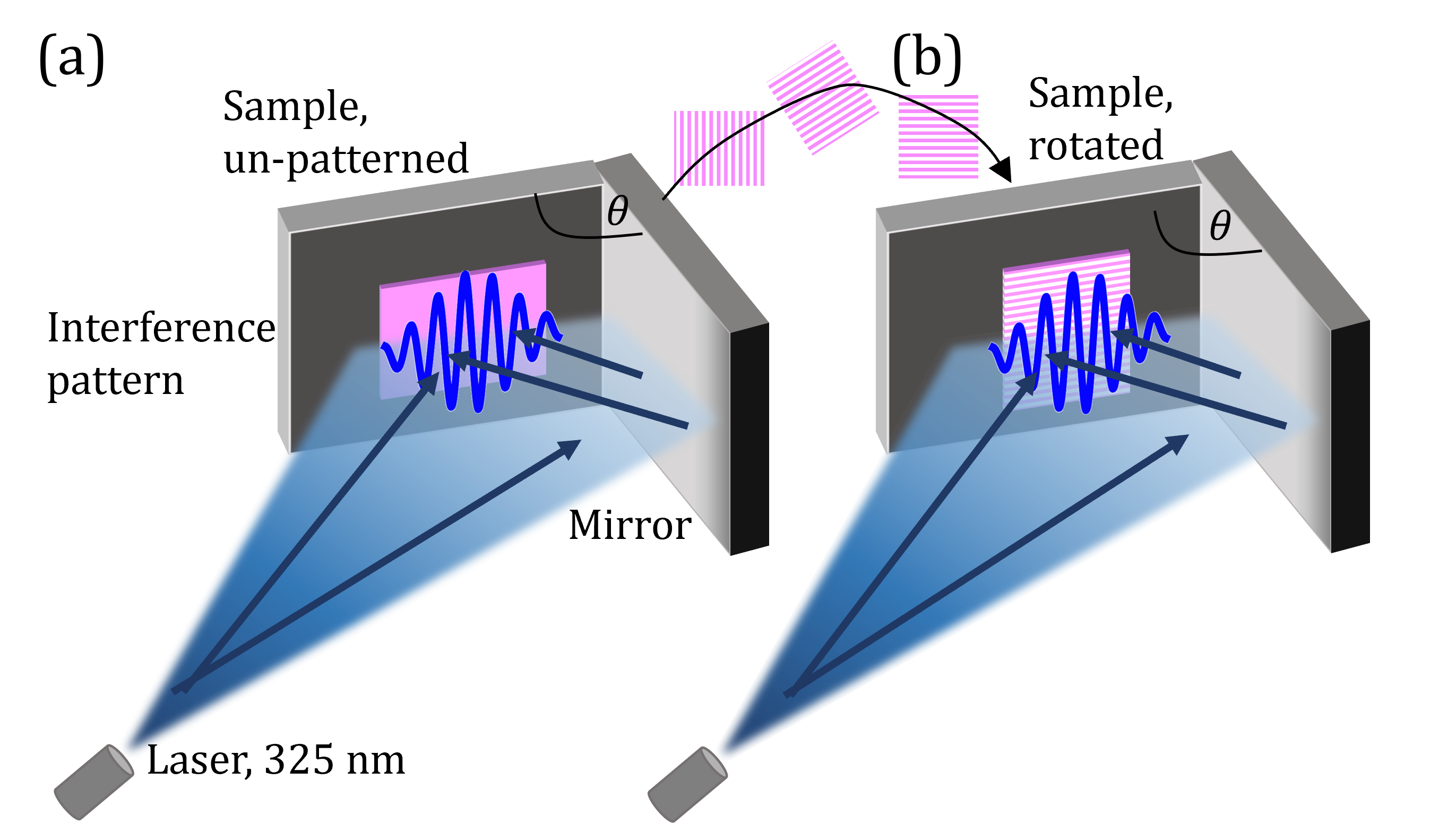}  \label{fig:inter_litho}
  \caption{\textbf{2D Interference Lithography schematic.} (a) A Lloyd mirror with angle $\theta = 11 ^\circ$ between the two mirrors is illuminated by a 325 nm laser field. The light from the laser and the light reflected from the mirror create an interference pattern on the sample. (b) After exposure, the sample is rotated by 90$^\circ$ and the process is repeated in order to generate the symmetric, large-scale 2D grating.}
\end{figure}

For scalable PhC fabrication, interference lithography was used. Figure \ref{fig:inter_litho} illustrates the core concept of the method. The SiN film was spin-coated with negative resist (NR7-250P, Futurrex) at 3000 rpm, baked at 150$^\circ$C for 60 s, and then exposed using a Lloyd’s mirror setup (325 nm wavelength). The sample was rotated 90$^\circ$ and exposed again for square lattice formation. It was then baked for 60 s at 100$^\circ$C, and developed with RD-61 (Futurrex) for 60 s. The pattern was transferred to the SiN layer using RIE, and NR7 was removed using He/O$_2$ RIE. This process produced a uniform 2D hole array over a 1 cm$^2$ area as depicted in Figure \ref{fig:last}(c). However, the Lloyd’s mirror setup imposed a limitation on the pattern's duty cycle: hole diameters smaller than 200 nm were challenging for our 800 nm pitch, though 400 nm diameters were readily attainable.

\subsubsection{Patterning of gold electrodes}
We spin-coated the GMR device with negative photoresist (nLOF 2020) at 4000 rpm, baked it at 110$^\circ$C for one minute, and patterned it using a Maskless Aligner MLA 150. After post-exposure baking at 110$^\circ$C, we developed the resist with AZ300 for 60s and deposited a 300 nm Au film using electron beam deposition. For lift-off, we vertically oriented the sample, submerged it in NMP, and placed it in a sonicator bath for about 10 minutes.

\subsubsection{CMOS backplane bonding}
We bonded the model JD2552 2K Voltage Drive Microdisplay from Jasper Display to the PhC using Cyclotene 3022-35 epoxy with a cold bonding method. After cleaning the PhC substrate with acetone, IPA, methanol, and a 1-minute O$_2$ plasma cleaning (for both substrates), we diluted Cyclotene in Mesitylene (2:1 ratio) and spin-coated the mixture on the PhC at 4000 rpm, forming a $\approx$ 400 nm thick film (confirmed through ellipsometery measurements). In a glove box, we baked the substrate at 150$^\circ$C for 10 minutes to evaporate Mesitylene and let it cool to room temperature. The bonding process was done in a glove box to avoid exposing the epoxy to oxygen. We then aligned and contacted the CMOS backplane with the substrate under a $\approx$ 200 kPa pressure using a custom bonding tool comprised of a heavy aluminum block, heated both to 150$^\circ$C at 7$^\circ$C/min, maintained this temperature for 10 minutes, then ramped to 280$^\circ$C at 1.5$^\circ$C/min. After 120 minutes at 280$^\circ$C and 200 kPa, we allowed the sample to cool to room temperature.    

\subsection{Numerical simulations}\label{subsubsec2_numerical}
We simulated the GMR with finite difference time domain (FDTD) software (Lumerical\textsuperscript{™} as well as Tidy3d \cite{hughes2021full}) using PML boundary conditions in the z-direction (out-of-plane) and periodic ones in the x and y (in-plane) directions. The simulations assume dispersionless materials, a 150 nm SiN hole radius, a 100 nm SiN thickness, a 500 nm LN thickness, and a 2400 nm SiO$_2$ film thickness. We assumed air and SiO$_2$ top and bottom claddings, respectively. The refractive indexes used were of $n_{SiN} = 1.98$ for the SiN and $n_o = 2.20748$ ($n_e = 2.14121$) for the ordinary (extraordinary) axis of the LN. To estimate the $Q$-factor versus PhC hole diameter, we used Legume, an open-source guided mode expansion (GME) software \cite{Minkov2020}. For the LN GMR modulator and bonded CMOS backplane, we used finite element method simulations in COMSOL\textsuperscript{™} to calculate the electric field distribution for a given electrode bias voltage.

\begin{figure}
  \centering
  \includegraphics[width=80mm]{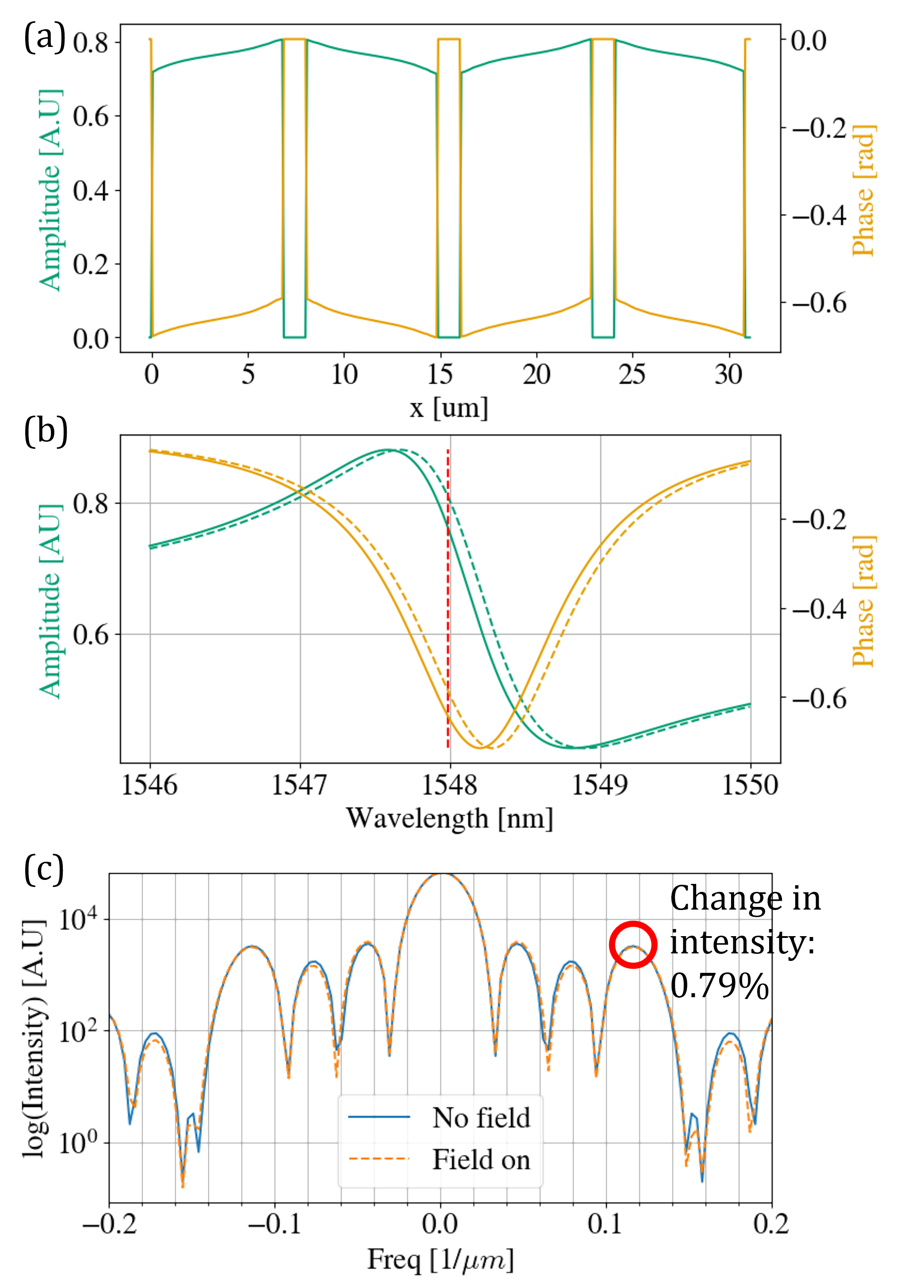}
  \caption{\textbf{Fourier analysis of the 1D LN GMR SLM.} (a) The cross section of the SLM profile, derived from the change in refractive index. (b) The phase (orange) and amplitude (blue) of the original (full line) and shifted (dotted line) reflection profiles. (c) the log of the intensity at the far field, derived from the Fourier transform of (a). The red circle denotes the first order. The change in intensity in the first order is 0.79$\%$.}
  \label{fig:theory}
\end{figure}

\subsubsection{Field overlap calculation}
We estimate the change in the first-order diffraction intensity from coupled mode theory \cite{haus1991coupled}. We work in the paraxial regime where the fields can be treated as scalar. We treat the cavity mode as a wave bouncing back and forth between two interfaces, and assume the cavity layer can be substituted for $Q/2\pi$. We assume these are reasonable assumptions allowing to apply these tools in this context. In this formalism, the goal of beamforming is to steer power from a fixed incident free-space mode with wave vector $\textbf{k}_{in}$ to a desired far-field mode possessing wave vector $\textbf{k}_{out}$. For simplicity, we consider only 1-dimensional beam steering along the $x$-axis. We assume that the GMR structure lies in the $xy$ plane. We denote the mode profiles of the input and output modes in the GMR structure by $e_{in}$ and $e_{out}$, respectively. Since LN is non-magnetic, the $x$ component of the wave vector of a given mode will be the same in free space and the LN - we denote the $x$-components for the two modes $k_{in}$ and $k_{out}$. 

The coupling between the input and output modes can be controlled by the permitivity perturbation, $\Delta \epsilon$, induced by the CMOS backplane. The resulting coupling coefficient is given by the overlap integral
\begin{equation}\label{eqn: Haus_Coupling}
    \kappa = \frac{1}{2} \frac{\int  e_{in}^* e_{out} \Delta \epsilon dV}{\int \epsilon \|e_{in}\|^2 dV}
\end{equation}
taken over a unit cell of the GMR structure \eqref{eqn: Haus_Coupling}. For a coupling constant $\kappa$, GMR quality factor $Q$, and normalized characteristic propagation distance $\frac{Q}{2 \pi}$, the relative intensity coupled to the output mode is given by \cite[Eq. 4.13]{haus1991coupled}
\begin{equation}
    \sin(\| \kappa \|\frac{Q}{2\pi } )
\end{equation}
Using the numerically simulated $\Delta \epsilon$ along with mode profiles corresponding to the wave numbers for the incident and targeted output modes, we can calculate the expected power steered by the CMOS backplane. Assuming a uniform $\Delta n = 10^{-4}$ and using our operation point ($\lambda_0 = 1548$ nm, $\eta = 0.9$, $Q=1400$, $E =2$V), we'd expect a change on the order of 0.7$\%$.

\subsubsection{Fourier analysis}
In order to evaluate the change in the far-field pattern as a function of the system parameters, we analyze the expected modulation by calculating the near field phase and amplitude SLM profile. Figure \ref{fig:last}(k) shows  the change in refractive index in the LN layer induced by the electric field simulated in COMSOL \textsuperscript{\textregistered} (Figure \ref{fig:last}(i)) using Equation \ref{eq:delta_n}. In order to get an upper bound on the modulation, we choose to take a cross section of this profile at the interface with the electrodes, where the field is largest. 
We calculate the change in resonant wavelength caused by the change in refractive index from Equation \ref{eq:delta_lambda}. The new (local) resonant frequency $\omega_0'$ and loss rate $\gamma'$ are then
\begin{equation}
    \begin{aligned}
    \omega_0' = \frac{2\pi c}{\lambda + 
\Delta \lambda}\\
    \gamma' = \frac{\omega_0'}{2Q}
    \end{aligned}
\label{eq_change}
\end{equation}
These values are used to calculate the  spatially dependant phase and amplitude phase and amplitude according to the Fano-shaped reflectivity profile \citep{elkabbash2021fano}:
\begin{equation}
R= \left\|a_1 +\frac{a_2}{\omega - \omega_0 + i\gamma}\right\|^2,
\label{eq_fano}
\end{equation}
where $\omega = \frac{2\pi c}{\lambda}$ is the frequency of the input light, $\lambda$ is the wavelength, $c$ is the speed of light, $\omega_0$ is the resonant frequency, $\gamma = \frac{\omega_0}{2Q} $ is the loss rate, $Q$ is the quality factor, $a_1$ is a constant accounting for the background and $a_2$ is the constant. In case we are simulating the GMR-LN, where the electrodes are directly written on the sample and do not reflect through the cross-polarized setup, we multiply by a binary profile that nulls out where the reflectance is zero (and can be obtained on the camera). In order to obtain the far field profile, the final profile is Fourier transformed. 

We now study the change in the first order intensity $\Delta I_1$. The first order is determined by the spatial frequency of the electrodes, and is located at $1/(W + G)$, where the width of the electrodes is $W=$6.8 $\upmu$m and the gap between the electrodes is $G=$1.2 $\upmu$m, yielding a spatial frequency of 0.125 1/$\upmu$m. We can monitor the change (in percent) of the first order intensity while varying the system parameters: $Q$, EO coefficient $r_{33}$, electric field strength $E$, refractive index $n$, and field overlap $\eta$. We combine the parameters affecting the change in refractive index, $\Delta n$, into a single parameter: $\Delta n / \Delta n_0$, where $\Delta n_0$ represents the value of $\Delta n$ obtained in our experimental configuration. Figure \ref{fig:theory} shows the results of this calculation for typical values that were fitted from Equation \ref{eq_fano} to the  measurement in our setup: with $\lambda_0 = 1548$ nm, $\eta = 0.9$, $Q=1400$, $a_2 = 206 2\pi$GHz and $a_1 = 0.61$. Figure \ref{fig:theory}(a) shows the phase and amplitude profile obtained in the near field of the SLM. Figure \ref{fig:theory}(b) shows the shift in the resonance profile, and Figure \ref{fig:theory}(c) shows the resulting far field in log scale, with $\Delta I_1 = 0.79\%$ at the first order.

\subsection{Experimental measurements}\label{subsubsec3}

\subsubsection{Optical measurements}
\begin{figure}
  \centering
  \includegraphics[width=84mm]{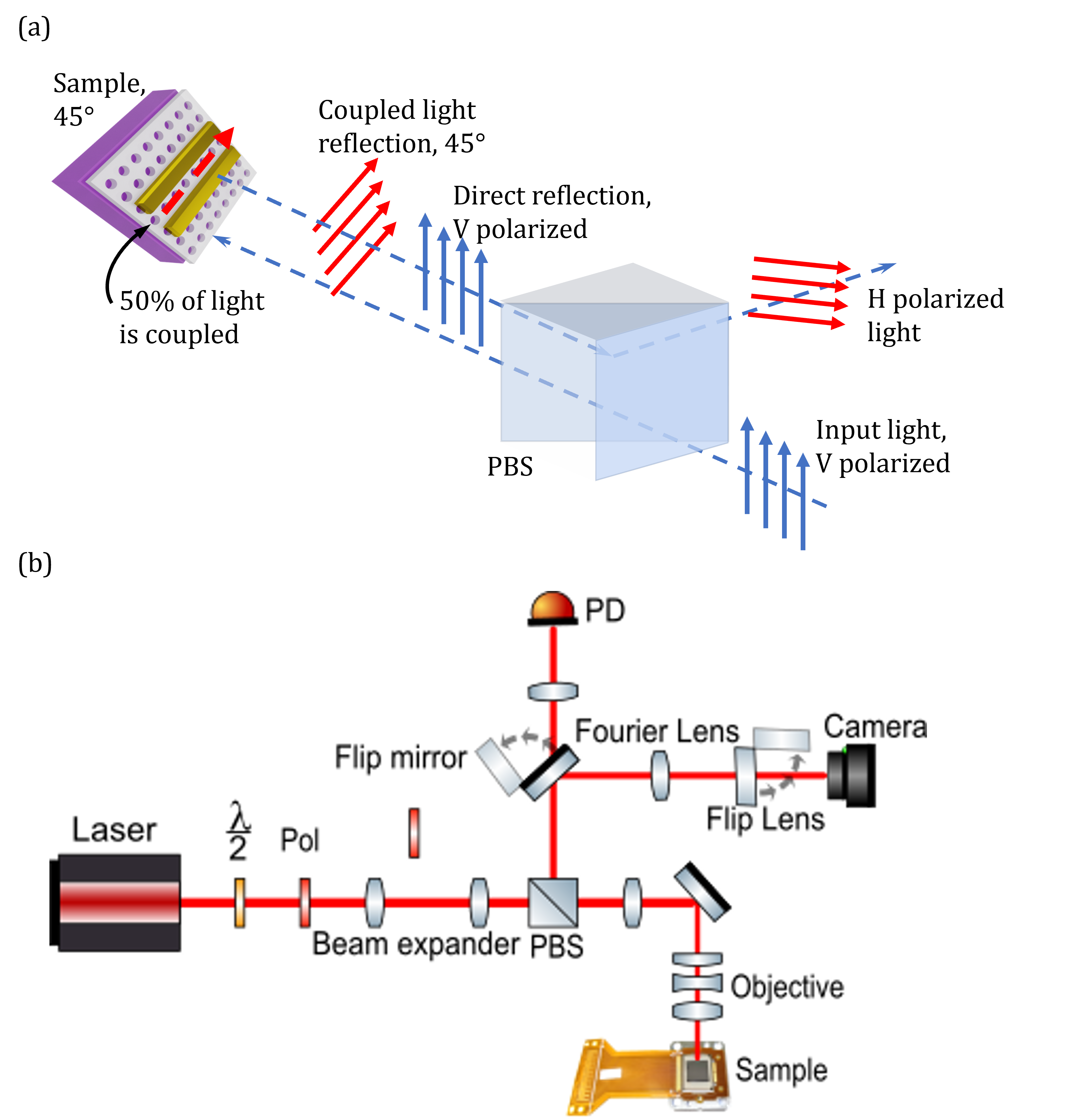}
  \caption{\textbf{Cross polarization optical measurement setup.} (a) Schematic of cross-polarization principal: The input light is V polarized, and passes through a PBS - polarizing beam-splitter. 50$\%$ of the light is transmitted to the sample, which is rotated at 45$^\circ$. The direct reflection polarization is left unchanged, but the coupled light is now polarized at 45$^\circ$. Upon returning to the PBS, the direct reflection is completely tramsitted, while 50$\%$ of the coupled light is reflected in H polarization. (b) Schematic of the optical setup. Pol - polarizer,  PD - photodetector. We have a flip mirror for moving between the camera and photodetector, and a flip lens to image the real plane or Fourier plane of the sample.}
  \label{fig:setup}
\end{figure}

We used a cross-polarized reflection imaging and spectroscopy technique to reduce background reflections as seen in Figure \ref{fig:setup}(a). The optical setup used for this experiment is schematically shown in Figure \ref{fig:setup}(b). We used a 15 mW CW  NIR tunable laser (Santec model 570) with a tuning range from 1480 nm to 1640 nm, running sweeps with intervals of 0.02 nm for measuring the resonances. The reflected power at each wavelength was measured on a VIS-IR camera (OWL 1280), which enabled us to integrate over the modulated areas in real space (or Fourier space). 
For our bandwidth measurements, we used a high-speed, temperature-compensated APD (Thorlabs APD450C InGAs), with a bandwidth of 0.3 - 1600 MHz. For some of our measurements, we added a spatial filter in the image plane to block the direct reflection. Our samples were mounted on a kinematic mirror mount to allow for direct vertical incidence. 

The measured GMR is fitted with a Fano lineshape from equation \ref{eq_fano}. The reason we expect this lineshape in our cross polarized setup is due to two contributions: (1) imperfections in the cross polarized setup, allowing for some direct reflection to pass and (2) reflection from the back side of the LN, where the silver paste and aluminum act as a mirror for the forward emission of the GMR. We measured the resonance of interest to have $Q \approx 1400$. This $Q$-factor can be further increased by reducing the hole size.

\subsubsection{DC measurements}\label{subsubsec2_DC}

For our DC tuning measurements, we mounted the sample on a custom designed chip carrier using thermally conductive silver paste. The sample was temperature controlled using a thermo-electric cooler (TEC). The electrodes were wire bonded to the chip carrier. We applied different voltages, ranging from 200 V to -200 V (at 10 V intervals) using a Keithley 2450 voltage source. At each voltage, we ran an automated wavelength sweep with a resolution of 0.02 nm, taking an image of the sample at each wavelength (camera settings are kept constant throughout the sweep). The images are then analyzed by integrating over the same desired area for the total reflection intensity at each wavelength. The final intensity vs. wavelength is fitted with a Fano line shape and the central wavelength is extracted. 

We perform the voltage sweep 6 times for each pixel to compute error bounds for each voltage. Sweeps were performed half of the time from -200 V to +200 V, and half of the time in reverse to avoid poling effects of the LN. 

\subsubsection{RF measurements}\label{subsubsec2_RF}
The RF measurements were conducted in our optical setup with the  Thorlabs APD450C photodetector (1.6 GHz 3-dB bandwidth) and a spectrum analyzer (Rigol RSA3030E) that can measure up to 3 GHz. The spectrum analyzer we used has an output port with an output of $ V_{rms} = 2$ V, that we used for the bandwidth measurement as the input driver to our 1D SLM. We removed the background noise by subtracting the measured signal when the laser is off. The measurement noise is primarily due to mechanical instability of the setup as well as high temperature sensitivity. To extract the bandwidth we fit the measured intensity data to a second order frequency response
\begin{equation}
I(s) =  \left(1 - \left(\frac{s}{s_0}\right)^2\right)^2  + \left(2\zeta\frac{s}{s_0}\right)^2 
\label{eq_complexPole}
\end{equation}
Where $s$ is the Laplace domain frequency ($f = \|s/2\pi i\|$), and $s_0$ and $\zeta$ are the natural frequency and damping factor for the second order system. The 3 db bandwidth was extracted from the fit. \\

\begin{figure}
  \centering
  \includegraphics[width=77mm]{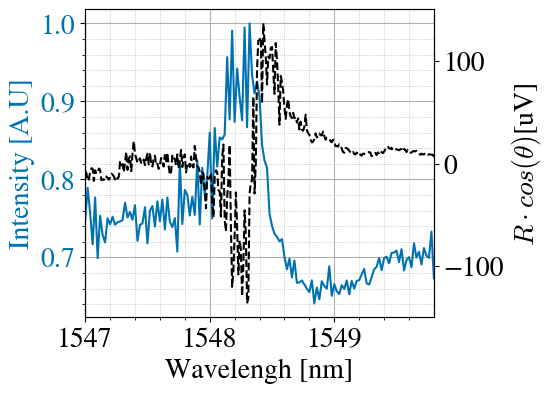}
  \caption{ \textbf{Lock-in measurement.} Blue shows the reflected intensity per wavelength. Dotted black shows the lock-in measurement $R\cdot \cos (\theta)$.}
  \label{fig:lockin}
\end{figure}

We used a lock-in amplifier (Stanford Instruments) to measure the small amplitude variation of the signal. We applied a 2V signal at 10 MHz to our device, and measured the lock-in reflection amplitude ($R$) and phase ($\theta$) while sweeping the laser wavelength. Figure \ref{fig:lockin} shows the results of this measurement. Comparing the lock-in measurement (black) with the measured GMR reflectance (blue), we see that $R\cdot \cos (\theta)$ is positive below resonance and negative above, with a zero crossing at the resonance peak. This is expected from a change coming from a resonance. 

\bmhead{Acknowledgments}

We thank the MITRE Quantum Moonshot Program for program oversight and funding. We thank the MIT.nano staff for fabrication assistance. S.T.-M. was supported by the Schmidt Postdoctoral Award and the Israeli Vatat Scholarship. C.L.P. was supported by the Hertz Foundation Elizabeth and Stephen Fantone Family Fellowship. The views and conclusions contained herein are those of the authors and should not be interpreted as necessarily representing the official policies or endorsements, either expressed or implied, of the United States Air Force, the Air Force Research Laboratory or the US Government.

\bmhead{Data availability}
The data that support the findings of this study are available from the corresponding author upon reasonable request.
\pagebreak

\begin{appendices}



\section{Three Pixel Theoretical Analysis}\label{secA1}%
\begin{figure}
  \centering
  \includegraphics[width=77mm]{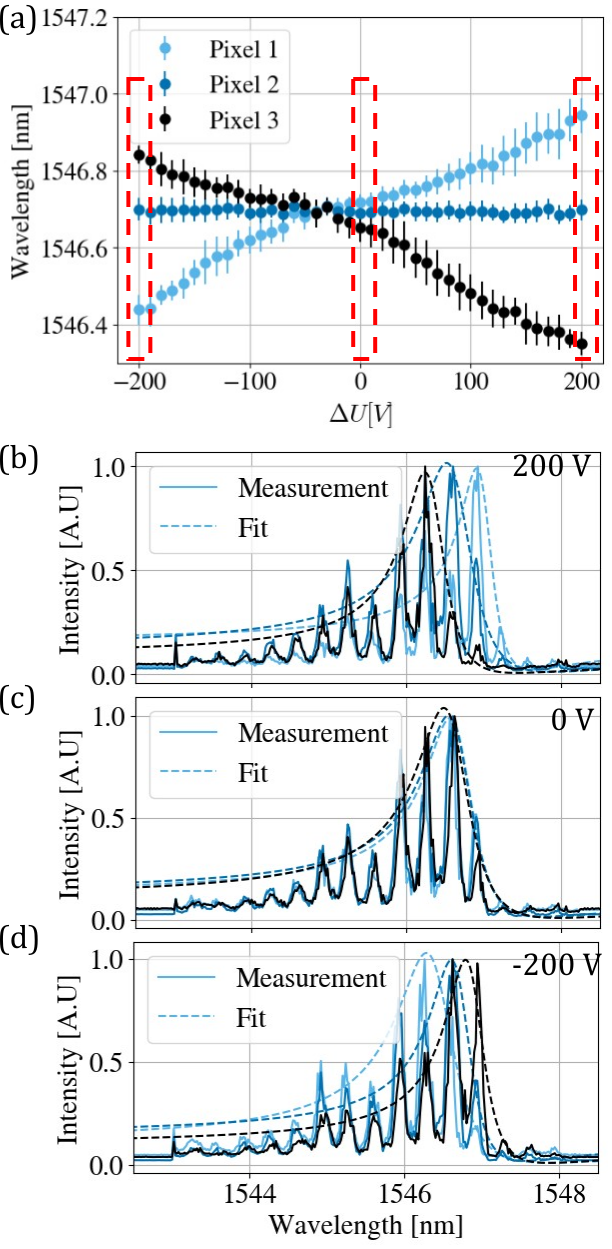}
  \caption{ \textbf{Measured resonant wavelength shift per pixel for a 3 pixel LN GMR SLM.} (a) Wavelength shift per for the three pixels: light blue, dark blue, and black. Red rectangles show the points presented in (b), (c) and (d). (b) Raw (full line) and fitted (dotted line) data for $\Delta U = 200 V$, (c) $\Delta U = 0 V$ and (d) $\Delta U = -200 V$.}
  \label{fig:shifts}
\end{figure}

Here we analyze the theoretical far field distribution for the three pixel SLM that we tested in Sec. \ref{subsubsec2_Charachterization}. We start by measuring the reflectivity profile of each pixel to account for possible differences due to angle or fabrication imperfections. Figure \ref{fig:shifts}(a) shows the wavelength shift per pixel in our setup for different applied voltages. Pixel 2 is the center pixel that is grounded on both sides, therefor has no potential difference across it while applying a voltage to the other pixels, and should therefore experience a minimal shift. The variance in this pixel gives us a measure of the stability of the measurement. Indeed the error for all measurements is of the same order. The error was calculated as the standard deviation over 6 measurements.

\begin{figure}
  \centering
  \includegraphics[width=77mm]{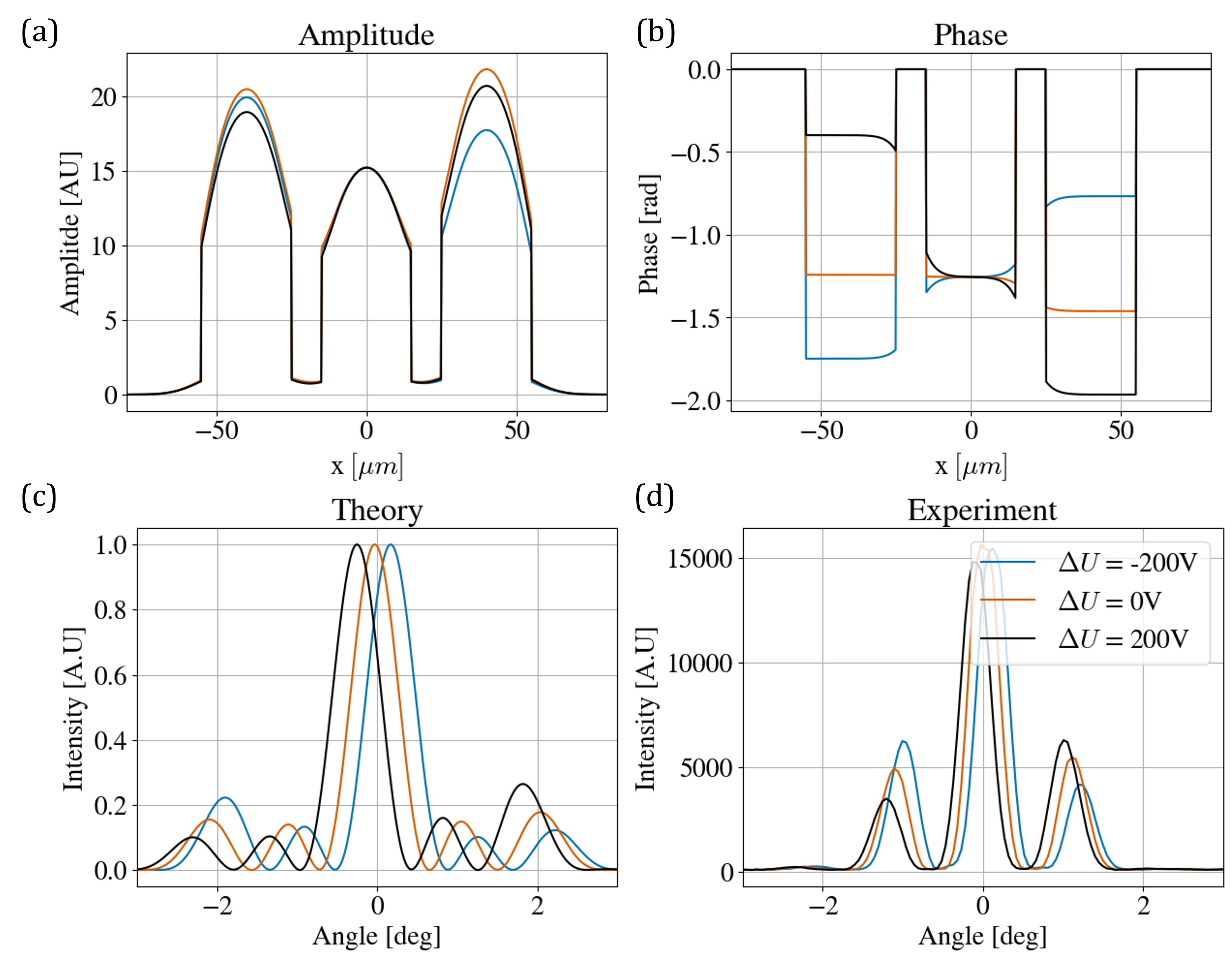}
  \caption{\textbf{Theoretical calculation of far field distribution using a 3 pixel LN GMR SLM.} (a) Amplitude and (b) phase of the theoretically calculated near field at $1547$ nm, for an applied voltage of -200 V (blue), 0 V (orange) and 5 V (black). (c) Resulting far field theoretical pattern and (d) experimental cross section at $1547$ nm for different applied voltages.}
  \label{fig:theory1}
\end{figure}

As can be seen clearly in Figure \ref{fig:shifts}(b-d), our measurements show that the resonance of interest has a higher frequency modulating it, corresponding to an etalon which comes from the insulating layer of the LN. We can filter out this frequency. We fit a Fano lineshape to the filtered measurement, and can estimate the phase and amplitude at each wavelength. Next, we build a digital twin of the three pixel SLM for each voltage, by taking these phase and amplitude value and tiling three 20 $\upmu$m sized pixels, separated by 10 $\upmu$m electrodes. In order to account for the fact the the near field image is not uniform, we convoluted each pixel with a Gaussian. Figure \ref{fig:theory1} shows the theoretical SLM near field and far field, alongside the measurement. In order to calibrate the measured far field angles, we put a known grating instead of the sample. Scattering from the grating caused the light to go through our cross polarized setup. For a grating with a distance $d$ between closest repeating points, diffraction peaks will appear at angles $\theta$ according to Bragg's law
\begin{equation}
n\lambda = 2d sin(\theta)
\label{bragg}
\end{equation}
So the far field angle for every x on the camera can be calculated directly by finding the distance between peaks $x_{peak}$ (in pixels)
\begin{equation}
\theta = \sin^{-1}(\lambda \nu) = \sin^{-1}\left(\frac{x\lambda}{2dx_{peak}}\right)
\label{farfieldAngle}
\end{equation}
In the measurement, higher orders are suppressed, but the trend is clear. Discrepancy between theory and measurement arises from the instability of the setup (shown by the error in the wavelength sweep), temperature variations as well as additional etalon which is unaccounted for.

\section{Phase space analysis}\label{secA2}%
\begin{figure}
  \centering
  \includegraphics[width=77mm]{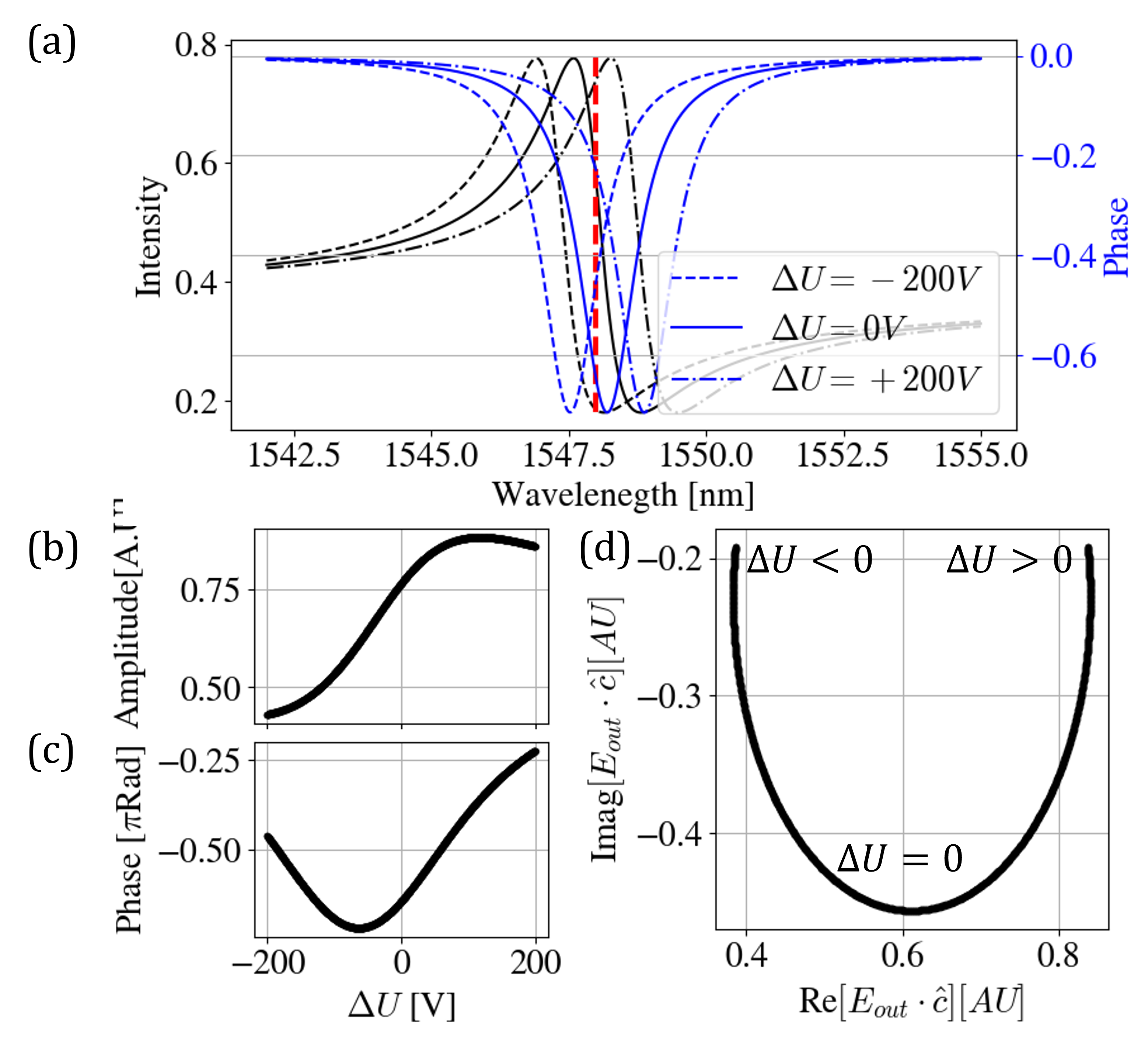}
  \caption{ \textbf{Phase space analysis.} (a) Fano shape with realistic parameters Intensity (black) and phase (blue), applied potential difference of $\Delta U$ = -200 V (dashed line), $\Delta U$ = 0 V (solid line) and $\Delta U$= 200 V (dash-dot line). Red shows our working wavelength. (b) Amplitude and (c) phase at our working point. (d) Real and Imaginary part of the field for varying voltage.}
  \label{fig:phasespace}
\end{figure}

The operation principle of the LNoS is that phase and amplitude changes are acquired by locally shifting a resonance. We can analyze the phase and amplitude accumulation in realistic conditions by looking at the phase space of the complex amplitude of the field at a given wavelength. 
We start by calculating the Fano shaped reflection, using  realistic values from our experiment: $Q = 1400$, $a_1 = 0.611$, $a_2 = 206.47$ rad/nsec and $\lambda_0 = 1547.9$ nm (see Equation \ref{eq_fano}), and $\eta = 0.9$. We examine the field in the three pixel arrangement, where the distance between pixels is $L$ = 30 $\upmu$m with a varying potential difference $\Delta U$ from -200 V to +200 V, and take the field to be $E=V/L$. Figure \ref{fig:phasespace} show the results of this analysis. It shows an extinction of $\sim$ 4 times in intensity. 

\section{Theoretical full stack performance}\label{secA3}%
\begin{figure}
  \centering
\includegraphics[width=77mm]{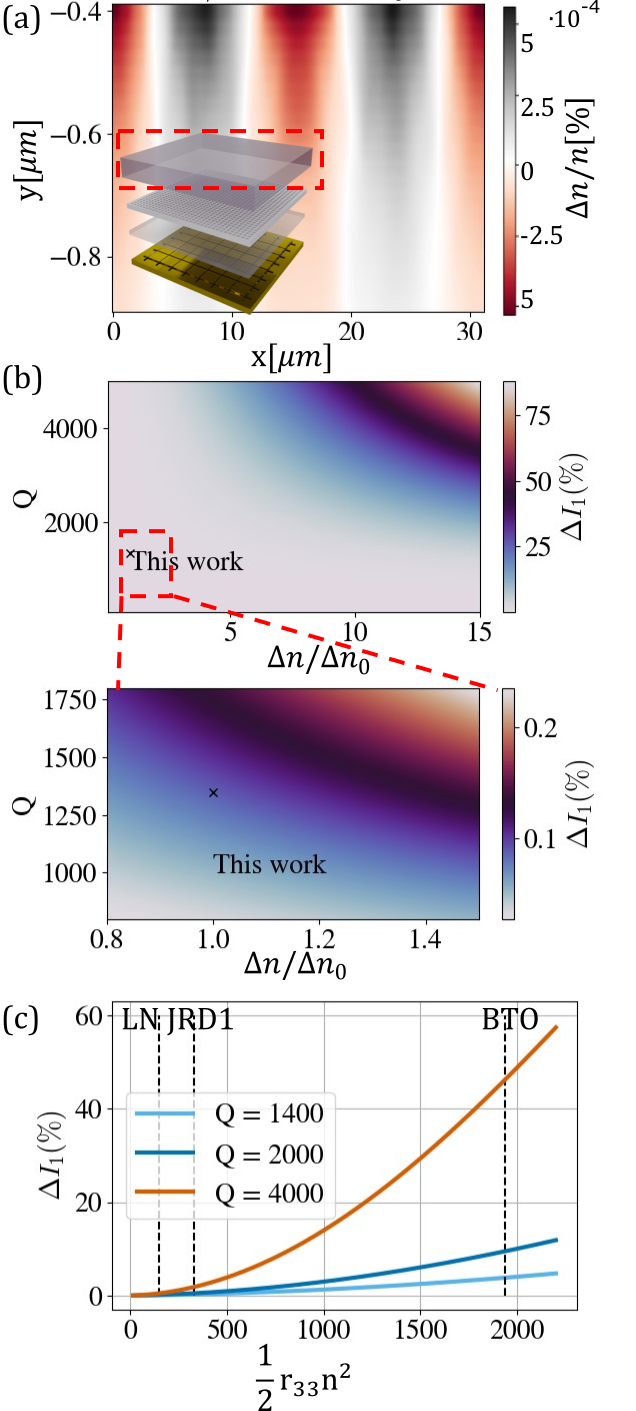}
  \caption{ \textbf{Performance including epoxy.} (a) Relative change in refractive index $
  \Delta n / n$ inside the LN layer, when simulating the entire stack with epoxy. (b) The change (in percent) in the first diffraction order's intensity $\Delta I_1$ for varying $Q$-factor and $\Delta n/ \Delta n_0$, where $\Delta n_0$ is the change in refractive index calculated for our experimental working point, arising from refractive index n, EO coefficient $r_{33}$ and electric field strength. Our work has $Q = 1400$ and $\Delta n/ \Delta n_0 = 1$, and is highlighted in the plot by $\times$. Bottom shows a zoomed in image of the area around our working point. (c) $\Delta I_1$ for a $5V$ peak bias voltage, $Q = 1400$ (light blue), $Q = 2000$ (dark blue) and  $Q = 4000$ (orange) as a function of the modified EO coefficient $\frac{1}{2}r_{33}n^2$ for LN (this work), JRD1, and BTO.}
  \label{fig:withepoxy}
\end{figure}

To account for the presence of a bonding layer between the CMOS backplane and GMR, we re-compute the Fourier analysis in Sec. \ref{subsec2}. Figure \ref{fig:withepoxy} shows the  results of this analysis for the stack including the epoxy layer. In this case, our work is calculated to have $\Delta I_1 \approx 0.088\%$. We show the materials depicted in Figure \ref{fig:last}(l): JRD1, LN and BTO. Also in this case, BTO would offer a significant advantage. It is clear that adding the epoxy layer requires higher Q factors for significant modulation. We consider our working $Q$-factor of 1400 and the higher $Q$-factor of 2000, but also examine an even higher $Q$-factor of 4000. A $Q$-factor of 4000 is achievable in the same settings with a hole diameter of $275$ nm, still easily fabricated using our large scale fabrication methods. Combing BTO with this high $Q$ will lead to an increase in the modulation strength up to $\sim 50 \%$. 

\end{appendices}


\bibliography{sn-bibliography}


\end{document}